# Erdős–Turán photonic Ising machines with record-high coupling resolution

Huaqiang Li[1†], Guangfeng Wang[1†], Erez Hasman[2], Bo Wang[1*], Xianfeng Chen[1,3,4]

[1]*State Key Laboratory of Photonics and Communications, School of Physics and Astronomy, Shanghai Jiao Tong University; Shanghai, 200240, China.*
[2]*Atomic-Scale Photonics Laboratory, Russell Berrie Nanotechnology Institute, and Helen Diller Quantum Center, Technion – Israel Institute of Technology; Haifa, 3200003, Israel.*
[3]*Shanghai Research Center for Quantum Sciences; Shanghai, 201315, China.*
[4]*Collaborative Innovation Center of Light Manipulations and Applications, Shandong Normal University; Jinan, 250358, China.*

[†]These authors contributed equally to this work.
*Corresponding author: wangbo89@sjtu.edu.cn (B.W.)

**One sentence summary**: By uniting *Sidon* set mathematics with momentum-space photonics, we overcome the long-standing resolution bottleneck in Ising machines, enabling unprecedented coupling precision and opening a scalable route to solving real-world optimization problems.

**Abstract:** Ising machines have emerged as promising platforms for efficiently tackling a wide range of combinatorial optimization problems relevant to resource allocation, statistical inference and deep learning, yet their practical utility is fundamentally constrained by the coarse resolution of spin-spin couplings ($J_{ij}$). Current implementations, relying on direct modulation of physical parameters, achieve at most 256 discrete coupling levels, which severely hinder the faithfully modeling of arbitrary real-valued interactions in realistic applications. Here we present a novel photonic Ising machine that encodes spins in random lattices while programming couplings in the momentum space of light. By introducing the *Sidon* set—a mathematical structure ensuring pairwise difference uniqueness— and employing the Erdős–Turán bound, we establish an optical framework in which each spin pair can be assigned a unique $J_{ij}$. This approach decouples the resolution limit from hardware modulation to the spatial precision in the momentum space of light. Experimentally, we demonstrate a record-high coupling resolution of 7,038 on a simple photonic platform, surpassing previous Ising machines. Our results highlight the power of uniting discrete mathematics with momentum-space photonics, paving the way toward scalable Ising machines capable of faithfully modeling real-world optimization problems.

## Introduction

Physical solvers of combinatorial optimization problems — particularly those are nondeterministic polynomial-time hard — have drawn increasing interest in recent years to serve as alternative computing paradigms beyond conventional von Neumann architectures. The Ising Hamiltonian, $H = \sum_{ij} J_{ij} s_i \cdot s_j$, is mathematically equivalent to quadratic unconstrained binary optimizations (QUBOs)[1,2]. It is also a universal representation encompassing many classes of combinatorial optimization tasks such as resource allocation, statistical inference and machine learning. Ising machines, hardware emulating such problems onto minimizing $H$, have been implemented across diverse physical platforms, including quantum systems[3–6], electronic circuits[7–9], surface acoustic waves[10], complementary metal-oxide-semiconductor (CMOS) oscillators[11,12], superparamagnetic tunnel junctions[13], and various photonics platforms[14–21]. For all these physical solvers, a core task of establishing an Ising machine is to use corresponding physical mechanisms to couple spin $i$ and spin $j$ with a specific coupling weight $J_{ij}$. The coupling resolution refers to the number of distinct values in the programmable coupling matrix $\{J_{ij}\}$, which determines how faithfully the hardware can represent the cost function of real-world problems. Therefore, coupling resolution is widely recognized as a critical parameter for accurate modeling of complex tasks[22,23]. For instance, in neuromorphic computing and artificial intelligence hardware accelerators[24], studies have shown that the resolution of weights (coupling) is a key factor affecting model performance[25]. To date, the highest coupling resolution achieved in Ising machines was 256 (8-bit) in degenerate optical parametric oscillators[26] and an analog optical computer[27]. Further increasing the coupling resolution in these systems necessities great technique breakthrough, typically at the cost of scalability and hardware overhead.

In the 1980s, the inherent parallelism and massive interconnection capability of free-space optics have been leveraged to implement Hopfield-type neural-inspired computation[28,29]. Furthermore, spatial photonic Ising machines (SPIMs) were introduced by employing spatial light modulators (SLMs) to encode spins into optical phases and compute the Ising Hamiltonian through free-space light propagation[19], with promising advantages in optical parallelism and therefore greatly reducing the computation consumption of electric components by free-space matrix operation. However, the inherent all-to-all connectivity of free-space light propagation also makes it challenging to construct arbitrary $J_{ij}$. In the past few years, many approaches have been proposed to improve the coupling matrix $\{J_{ij}\}$, using matrix transform[30–32], wavelength multiplexing, focal plane division[34], **k**-space modulation[35,36], as well as spin-product encoding [37]. The maximum coupling resolution reached in SPIMs is still 256 levels[27], which is constrained by the 8-bit phase modulation depth of the SLM. Mathematically, for a two-body spin model with $N$ spins, the maximum spin coupling resolution should be $C_N^2 = N(N-1)/2$. This limit, as a crucial criterion to simulate real-world combinatorial optimization problems, has been elusive in these Ising machines.

In this work, we establish an Erdős–Turán (ET) Ising machine that is able to precisely construct any coupling weight $J_{ij}$ to solve a wide range of Ising problems and arbitrary QUBO problems. We encode $N$ spins on an $L \times L$ square lattice of an SLM such that the displacement vector $r_{ij}$ between any pair of spins is unique, thereby naturally forming a *Sidon* set $\{\mathbf{r}_{ij}\}$ in $\mathbb{Z}^2$ [38]. According to the ET bound[39], the maximum number of arbitrarily programmable spins scales linearly with the lattice size, i.e., $\max\{N\} \sim \mathcal{O}(L)$, and the coupling resolution reaches the mathematical limit $C_N^2$ within the ET bound. This mathematical upper limit defines the ultimate modeling accuracy — the finest resolution of the coupling — for any two-dimensional computing architecture whose coupling weights are parameterized by spatial separations or phase gradients. This geometric conclusion is independent of the physical platform, being dictated solely by the dimensionality of the space that is used to encode the spin–spin couplings. Beyond the ET bound, partial constraints are imposed on the coupling weights by allowing repeated displacement vectors, thereby enabling a systematic scaling of the spins $N$ up to $L^2$, with the coupling resolution reaching $2L(L-1)$. Experimentally, we solved representative instances of the graph partitioning problem using ET SPIM, demonstrating its advantages in coupling precision, graph connectivity, and spin scalability, with a success probability comparable to that of simulated annealing implemented (SA) on a digital computer[40].

**The working principle of ET SPIM.**

A *Sidon* set is defined by the property of difference uniqueness, meaning that all pairwise differences between its elements are distinct. This structure, originally studied in additive combinatorics[38], is a powerful tool in coding and sequence design[41,42], sparse signal processing[43], harmonic analysis[44] and array geometry[45]. In our ET SPIM, *Sidon* set in $\mathbb{Z}^2$ ensures mutual orthogonal mappings of spin couplings in the **k**-space (S2.1 for details), enabling arbitrarily programmable spin models.

A simplified example of six-spin fully connected graph with unique couplings $J_{ij}$ is shown in Fig. 1a, where the edge colors represent the assigned coupling weights. The corresponding coupling matrix $\{J_{ij}\}$ is displayed in Fig. 1d, and all diagonal self-couplings are set to $J_{ii} = 0$. As a first step, we map all spins onto six sites (the blue pixels) on an $L \times L$ SLM (Fig. 1b) forming a *Sidon* set, which requires that the relative displacements between any two sites $\mathbf{r}_{ij} = (x_i - x_j, y_i - y_j)$ are unique. Under the condition of centrosymmetric ($\mathbf{r}_{ij} = \mathbf{r}_{ji}$), the sampling is realized through random search algorithms (S2.2 for details). When a planewave laser illuminates the SLM (Fig. 1c), optical phases are modulated on the blue pixels to as spin configuration $\{\theta_i\} = \{\theta_1, \ldots, \theta_N\}$, with $\theta_i \in \{0, \pi\}$. To prevent the background pixels from interfering with our designed spin model, we encode the background blank pixels with an antiferromagnetic (AM) phase distribution with a different super-pixel factor (S4 for details). In addition, we map the coupling weights $J_{ij}$ on the displacement space $\{\mathbf{r}_{ij}\}$, which is represented as a $(2L – 1) \times (2L – 1)$ grid (Fig. 1e). *Sidon* set guarantees that each $J_{ij}$ is independently assigned to a unique site in the $\mathbf{r}_{ij}$ space, yielding $J_{ij} = J(\mathbf{r}_{ij})$. The coupling weights $J_{ij}$ may take any discrete values—positive, negative, or zero—following an arbitrary distribution. Next, we generalize the Plancherel theorem of light to describe the conservation of the Ising Hamiltonian in random lattices between real space and **k**-space (S1 for details), that is,

$$H_\mathbf{k} = \iint_{-\frac{2\pi}{\Lambda}}^{\frac{2\pi}{\Lambda}} I(k_x, k_y) V(k_x, k_y) dk_x dk_y \propto \sum_{i,j} J(\mathbf{r}_{ij}) s_i \cdot s_j \tag{1}$$

The Ising Hamiltonian $H_\mathbf{k}$ can be interpreted as a Frobenius inner product between $I(\mathbf{k})$ and $V(\mathbf{k})$ in the **k** space. The modulation function $V(\mathbf{k}) = \sum_{ij} J(\mathbf{r}_{ij}) e^{i\mathbf{k} \cdot \mathbf{r}_{ij}}$ is obtained by Fourier transform (*FT*) of $\{J(\mathbf{r}_{ij})\}$ (Fig. 1g). We show that, with a proper designed resolution, we need only a 3-bit-depth of modulation for $V(\mathbf{k})$ to achieve the coupling resolution of 16 bit (S3 for details). This approach shifts the resolution limit from hardware modulation to spatial precision in **k**-space, and significantly improves the programmable coupling resolution of Ising machines. The intensity distribution of light $I(\mathbf{k})$ is captured by the camera (Fig. 1f), corresponding to the Fourier power spectrum of the modulated light field (Fig. 1c). The computational flow of the ET SPIM is illustrated in Fig. 1h. For a given problem instance, $V(\mathbf{k})$ is calculated at initialization and reused throughout the entire annealing process. The Hamiltonian $H_{\mathbf{k}}$ is evaluated through the Frobenius inner product of the measured intensity distribution $I(\mathbf{k})$ and the modulation function $V(\mathbf{k})$. This process leverages the inherent parallelism and near-zero-latency summation of spatial light propagation, enabling efficient optical acceleration.

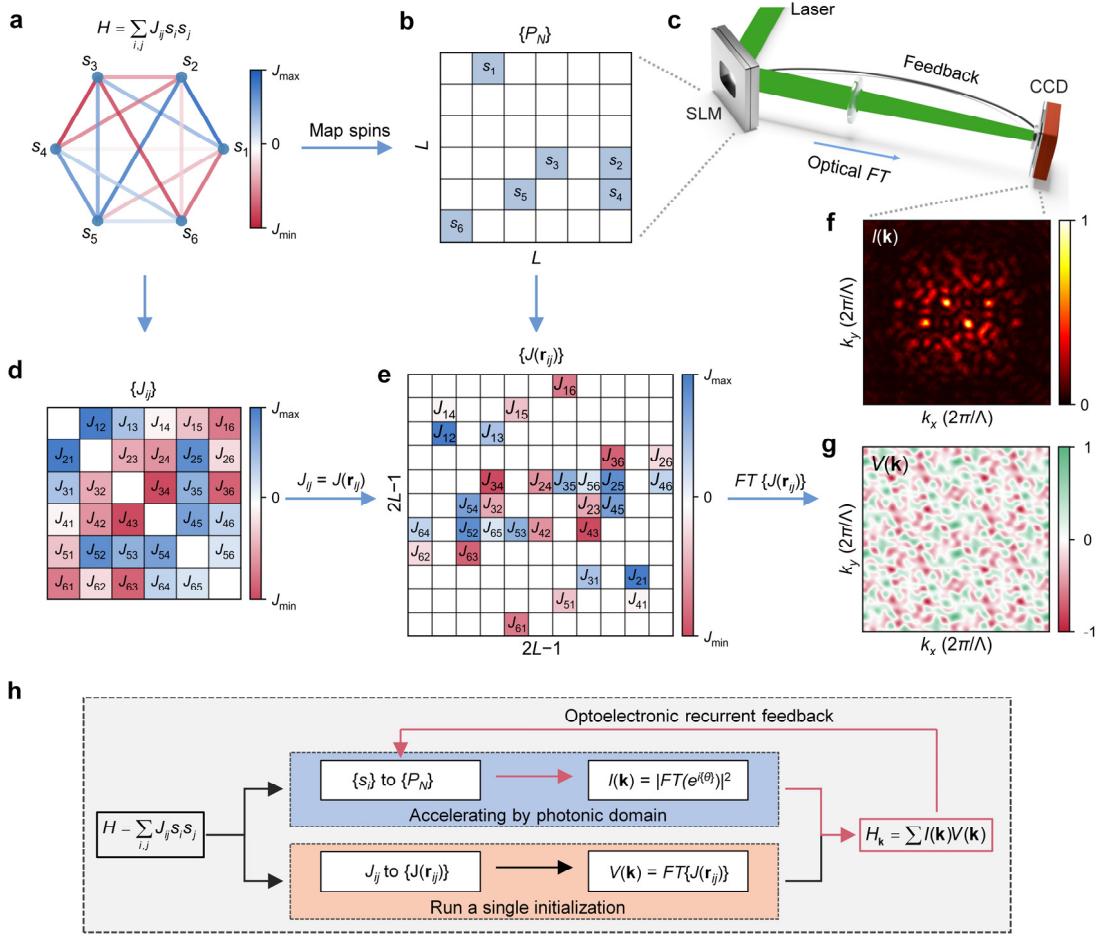

**Fig. 1 | Working principle and computational flow of the ET SPIM. a**, An example of a fully connected six-spin graph with unique couplings, where edge colors denote the weights $J_{ij}$. **b**, Spins are mapped onto six sites of a *Sidon* set $\{P_N\}$, each site is represented by an SLM super-pixel, with its phase modulated to 0 or π to encode the spin state. **c**, Schematic of the ET SPIM hardware, including an SLM, Fourier transform lens, and a camera for measuring intensity distribution. **d**, The adjacency matrix $\{J_{ij}\}$ corresponding to the graph in (**a**), off-diagonal elements $J_{ij}$ and $J_{ji}$ are highlighted with the same color to indicate identical weights. **e**, Coupling matrix $\{J_{ij}\}$ mapped onto displacement space $\mathbf{r}_{ij}$ according to $J_{ij} = J(\mathbf{r}_{ij})$. The $\mathbf{r}_{ij}$ space is represented as a $(2L − 1) \times (2L − 1)$ grid. **f**, Fourier intensity distribution $I(\mathbf{k})$ in the **k**-space. **g**, Modulation function $V(\mathbf{k})$ obtained via discrete Fourier transform of $\{J(\mathbf{r}_{ij})\}$. **h**, computational flow of the ET SPIM: Frobenius inner product of $I(\mathbf{k})$ and $V(\mathbf{k})$ yields the total experimental energy $H_{\mathbf{k}}$, with spin configuration $\{\theta\}$ are iteratively updated by SA to approach the ground state.

## The SPIM's $J_{ij}$ programmability across ET bound.

Fig. 2a shows the phase diagram of SPIM's $J_{ij}$ programmability, regarding on the maximum distinct $J_{ij}$ achievable for different spin number $N$ that is encoded on a fixed $L \times L$ lattice. All discussions are based on all-to-all connections. This phase diagram is a general result adapting to all SPIMs utilizing a two-dimensional (2D) optical field and single Fourie transform. There are four typical regions, and only region A and B are achievable in SPIMs. The dashed black curve sets the $C_N^2$ limit for two-body spin models, and region D is unattainable from two-body spin models. Particularly, region A is within the ET bound ($\sim \mathcal{O}(L)$). There, the number of spin $N$ scales with the lattice size $L$, as $N \leqslant \mathcal{O}(L)$. In this case, it is always possible to find $N$ spins to form a *Sidon* set, such that all pairwise displacement vectors $\{\mathbf{r}_{ij}\}$ are distinct, yielding $N(N-1)/2$ (the blue curve) fully and independently programmable $J_{ij}$. The left panel of Fig. 2b shows an example of *Sidon* set (bule point) in region A and the corresponded $\mathbf{r}_{ij}$ counts (right panel). In region B, namely, beyond the ET bound, repeated $\mathbf{r}_{ij}$ will always occur for more than one spin pair. Therefore, even if the number of unique $J_{ij}$ is increasing (the red dots in Fig. 2a), it remains below the $C_N^2$ limit (the dashed black curve in Fig. 2a). To ensure physical implementability and preserve the closed-form convolutional structure in **k**-space, any repeated displacement vectors must share identical coupling weights, that is, whenever $\mathbf{r}_{ij} = \mathbf{r}_{mn}$, it follows that $J_{ij} = J_{mn}$. This trade-off facilitates the continued growth of both the number of spin $N$ and unique $J_{ij}$. This scaling law of spin number and coupling $J_{ij}$ naturally corresponds to the generative dynamics observed in many real-world complex networks[46]. Notably, such repeated $\mathbf{r}_{ij}$ enhance the robustness of ET SPIM against noise and stabilize its optical readout. When the 2D lattice is fully filled with $N = L^2$ spins, the unique $J_{ij}$ number reaches its geometric limit of $2L(L-1)$ (indicated by the yellow dashed line, Fig. 2a).

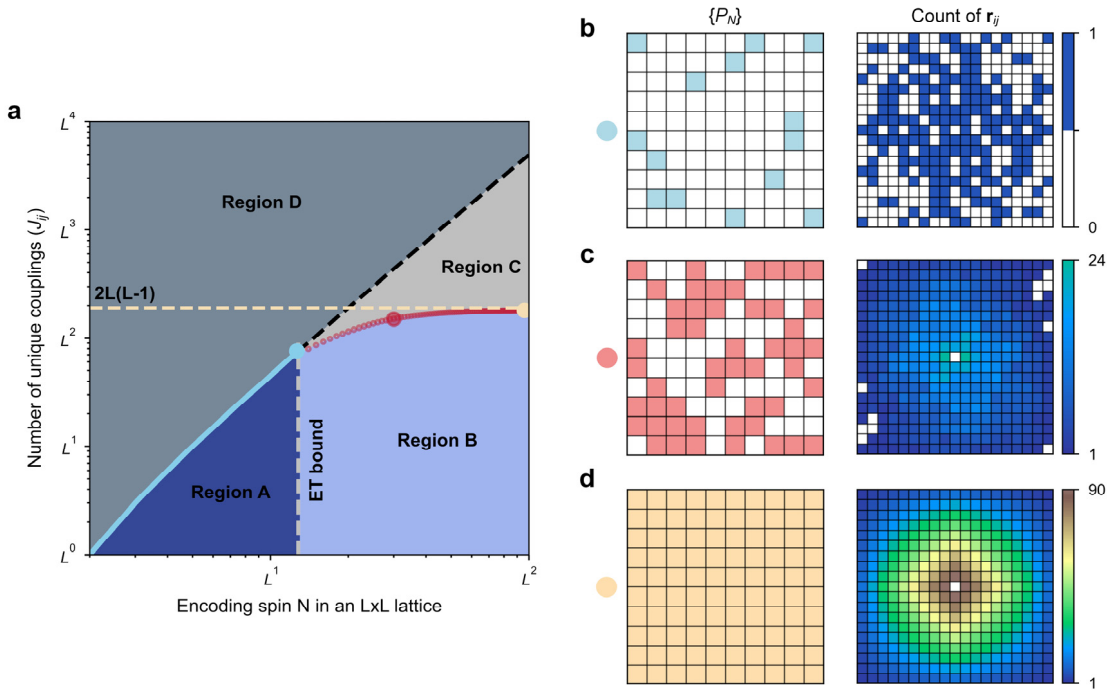

**Fig. 2 | The phase diagram of SPIM's $J_{ij}$ programmability in $\mathbb{Z}^2$. a**, The phase diagram of number of unique $J_{ij}$ as a function of spin $N$ encoded in an $L \times L$ lattice. The black dashed line denoted the limit $C_N^2$. ET bound is $N \sim \mathcal{O}(L)$. All discussions are based

on all-to-all connections. Region A: encodable with arbitrary $J_{ij}$; Region B: encodable with all-to-all but restricted $J_{ij}$; Region C: unattainable from the SPIM with a single Fourier transform; Region D: beyond two-body spin Hamiltonian; **b**–**d** left panel spin-site configurations $\{P_N\}$ corresponding to the three representative points (blue, red, yellow) marked in (**a**). Right: corresponding number of displacement vectors in the $\mathbf{r}_{ij}$ space, where color bars indicate the counts of $\{\mathbf{r}_{ij}\}$. **b**, Within the ET bound, spin sites form a *Sidon* set, where every displacement vector $\mathbf{r}_{ij}$ is unique, allowing all couplings $J_{ij}$ to be independently programmable. **c**, Above the ET bound, repeated $\mathbf{r}_{ij}$ gradually appear, while both the spin number and the unique $J_{ij}$ continue to grow. **d**, In the fully filled case with $N = L^2$, the number of distinct $\mathbf{r}_{ij}$ and $J_{ij}$ saturates at the geometric maximum $2L(L-1)$.

**Benchmark of ET SPIM in graph's connectivity.**

We experimentally evaluated the performance of our ET SPIM on Max-cut problems under varying graph densities d, where d is defined as the fraction of non-zero entries in the coupling matrix $\{J_{ij}\}$. *Sidon* set from a 10 × 10 lattice yields $N = 14$ spins (Fig. 2b, left), with each spin site encoded on the SLM as a super-pixel consisting 25 × 25 pixels. On these spin sites, we construct three representative graph instances with d = 0.2, 0.6, and 1. The corresponding coupling matrix $\{J_{ij}\}$ contain 18, 54, and 91 unique non-zero integers, randomly selected without repetition from the range [−46, 46]. The edge to vertex ($J_{ij} \rightarrow s_i s_j$) connections are illustrated in the insets of Fig. 3c-e. Notably, the spatial distributions of $V(\mathbf{k})$ (Fig. 3a) do not become more complex or fragmented for higher graph densities (S5 for details). This indicates that neither computational consumption nor precision is compromised as the number of unique $J_{ij}$ increases. For each graph density d, we solved 30 experiments on ET SPIM and 1000 simulations on a digital computer, with each run initialized from a random spin configuration $\{\theta\}$. The SA hyperparameters were optimized prior to solving each graph instance and were kept identical for both experiments and simulations[47]. Convergence of Ising energy was typically achieved within approximately 100 iterations. We adopted a success probability ($P_{suc}$) metric, which has been widely used in previous studies[16,48], to define success as achieving an Ising energy within 95% of the best solution obtained from the 1000 simulations[8].

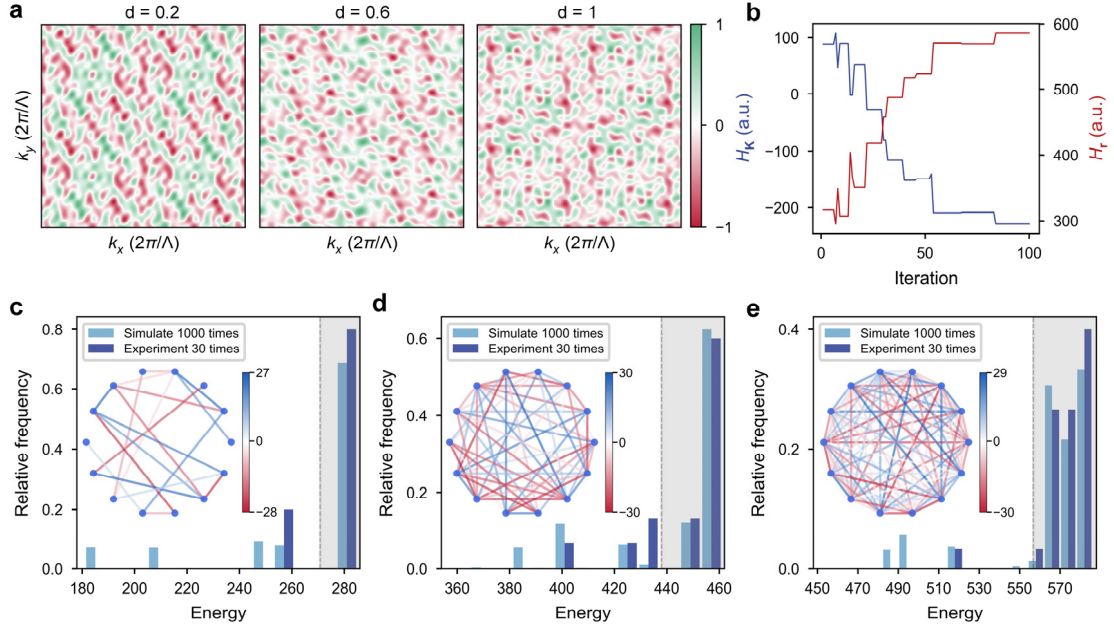

**Fig. 3 | Benchmarking of graph connectivity. a**, Three modulation functions $V(\mathbf{k})$ corresponding to graph densities d = 0.2, 0.6, and 1. **b**, Comparison of energy landscapes between the theoretical Hamiltonian $H_\mathbf{r}$ (red) and the experimental Hamiltonian $H_\mathbf{k}$ over 100 phase iterations. **c–e**, Ground-state energy distributions obtained from 1000 numerical simulations (light blue) and 30 experimental runs (dark blue). Insets show the corresponding connectivity graphs, with the edge colors representing the coupling weights.

Fig. 3c to e show bar plots of the ground-state energy distributions from simulations (blue) and experiments (dark blue). The experimental $P_{suc}$ values for the three graph densities — 80%, 73%, and 93% — closely match the simulation results of 68%, 77%, and 86%, with minor differences attributed to limited statistics. During the experiments, we also recorded the theoretical Ising Hamiltonian $H_\mathbf{r}$ at each iterative phase configuration to evaluate the accuracy of the experimental Hamiltonian $H_\mathbf{k}$. As shown in Fig. 3b, the energy landscapes obtained from $H_\mathbf{r}$ (red line) and $H_\mathbf{k}$ (blue line) exhibit near-perfect agreement across the entire iterative process, indicating that our approach faithfully represents the intended Ising energy profile. These results demonstrate that our system supports arbitrary graph connectivity with high fidelity, thereby confirming its flexibility in implementing both sparse and dense spin interactions.

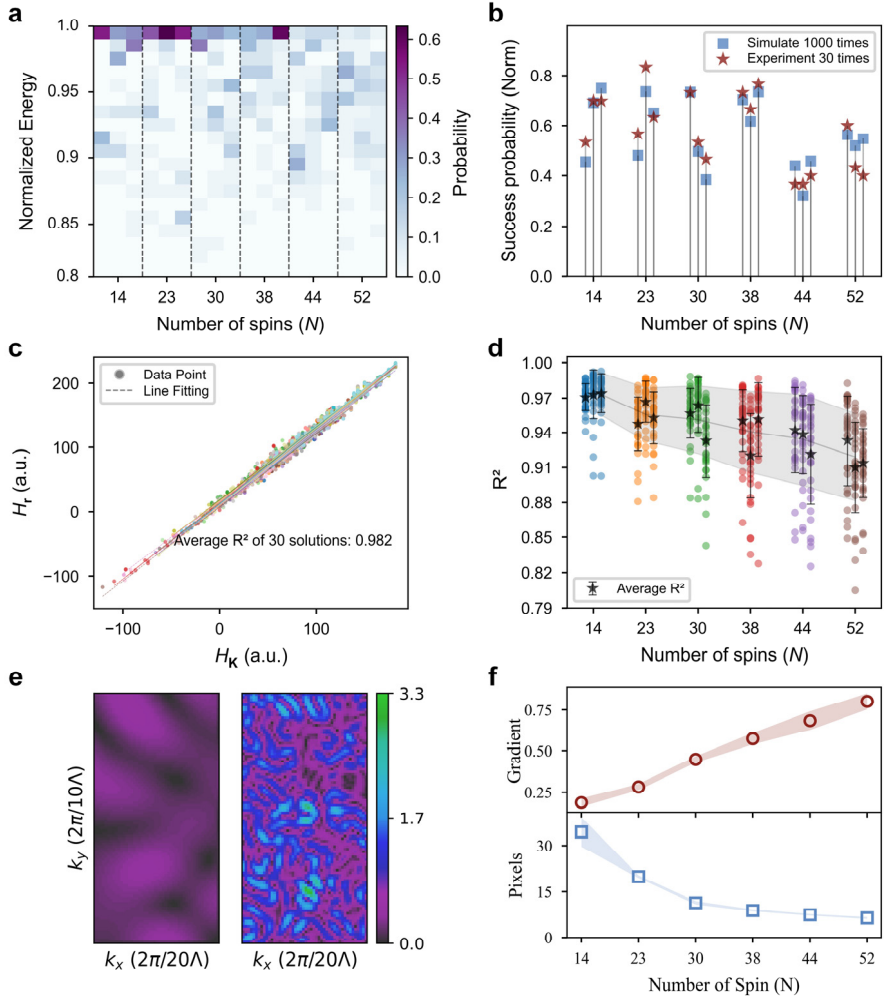

**Fig. 4 | Scalability and experimental performance of the SPIM within the ET bound. a**, Normalized ground-state Ising energy distributions from 30 experimental runs at different problem sizes. The color bar indicates the probability density. **b**, Success probability $P_{suc}$ compared between 30 experimental runs (red stars) and 1000 simulations (blue squares). **c**, Average coefficient of determination $R^2$ across 30 experimental runs at $N = 14$. The fitting line is obtained by least squares regression. **d**, Statistical distribution of $R^2$ for all six problem sizes. Black stars denote the average $R^2$. **e**, Local gradient magnitude maps of $V(\mathbf{k})$ for $N = 14$ (left) and (right) $N = 52$, covering 48×96 pixels. The color bar indicates the gradient magnitude. **f**, Average gradient magnitude (top) and boundary thickness of $V(\mathbf{k})$ (bottom) as a function of spin number.

**The scalability of ET SPIM within the ET bound.**

To investigate the scalability of the SPIM within ET bound, we increased the lattice size $L \times L$ to accommodate spin configurations of larger *Sidon* set. Specifically, we tested different spin numbers $N = 14, 22, 30, 38, 44$, and $52$, corresponding to lattice scales ranging from $L \times L = 10 \times 10$ to $60 \times 60$. To maintain proper encoding, the SLM super-pixels were adapted to the lattice scale and the extent of **k**-space. For each spin size, we constructed three fully connected graph instances (d = 1) with difference coupling matrix $\{J_{ij}\}$. Each $\{J_{ij}\}$ contains $N(N-1)/2$ non-zero integers, with are randomly selected from the range [−30, 30]. Similar to the previous section, we conducted 30 experiments and 1000 simulations for each instance. The ground state energies were normalized with respect to the best solutions obtained from the 1000 simulations. Fig.

4a shows the distribution of the normalized Ising energies obtained from the 30 experiments, with the color bar indicating the probability density. As shown in Fig. 4b, the $P_{suc}$ of the experimental results (the red stars) and simulation results (the blue squares) exhibit strong agreement across all tested instances, showing that the performance of the ET SPIM maintains consistent as the spin size increases. We performed a linear regression (ordinary least squares) between the $H_\mathbf{k}$ and $H_\mathbf{r}$, and evaluated the goodness of fit using the coefficient of determination ($R^2$). Fig. 4c shows the average $R^2$ that are obtained from the 30 experiments for spin number $N = 14$, yielding an average $R^2 = 0.982$. Fig. 4d summarizes the $R^2$ values across all the six tested spin numbers, with black stars representing the average $R^2$ for each instance. Although the average $R^2$ decreases to 0.91 at $N = 52$, this deviation remains within the system's tolerance range for spin state acceptance, as confirmed by the comparison of $P_{suc}$.

As the spin number increases, the structure of $V(\mathbf{k})$ and $I(\mathbf{k})$ become more intricate and fine-grained (S5 for details). This is because we must compress more information into a fixed pixel area, consequently increasing its susceptibility to experimental noise. We applied *Sobel operator*[49] to evaluate these variations of $V(\mathbf{k})$. Fig. 4e compares the local gradient magnitude maps that extracted from $V(\mathbf{k})$ for spin $N = 14$ (left) and (right) $N = 52$, with the latter exhibiting more intricate structures and steeper transitions at domain boundaries. We further adopted a geometric method that is commonly used to analyze structural transitions in cell membranes and phase-field models[50], to estimate the boundary thickness defined as the ratio of the interfacial area to the total contour length. As shown in Fig. 4f, the average gradient magnitude (top panel) increases approximately linearly with the spin number, while the boundary thickness (bottom panel) decreases from ~30 pixels for $N = 14$ to ~6 pixels for $N = 52$. According to the energy calculation of ET SPIM, $H_\mathbf{k} = \sum I(\mathbf{k})V(\mathbf{k})$, this makes an accurate pixel-level alignment between the $V(\mathbf{k})$ and $I(\mathbf{k})$ critical. As $N$ increases, the accurate alignment becomes more challenging, as any microscale mechanical vibration in the setup will cause $\mathbf{k}$-space misalignment and induce computing errors. Nevertheless, this issue can be suppressed by employing an ultra-stable platform with fine feedback control. The ultimate mathematical constraint arises from the finite pixel size of the SLM. According to the ET bound, the maximum number of spins (with arbitrary spin model) is limited to $N \sim 1000$ for a 1080 × 1080 pixels SLM.

**The scalability of ET SPIM above ET bound.**

Here we explore the spin model beyond the ET bound. Starting from *Sidon* set with $N = 14$ (Fig. 2b, left), and we gradually fill the spins to $N = 25, 50, 75$ and $100$ within a 10×10 lattice. For $N = 100$, the maximum number of unique $J_{ij}$ is $2L(L-1) = 180$. We constructed the graph density d = 1 for each instance (Fig. 5a), where each $J_{ij}$ is uniquely assigned an integer value from the interval $[-90, 90]$. As shown in Fig. 5b, ordered ground states emerge as the spin number $N$ increases, even though the $J_{ij}$ values are assigned randomly. This comes from the fact that repeated $\mathbf{r}_{ij}$ occur far more frequently for short range interactions (Fig. 2d, right panel), and this system effectively behaves similar to $J_1$-$J_2$ spin model in condensed matter physics with short-range interactions dominating the formation of ordered ground states[51]. Accordingly, the distributions of optical intensities $I(\mathbf{k})$ exhibit increasingly regular structures and progressively concentrate into narrower spots in $\mathbf{k}$-space.

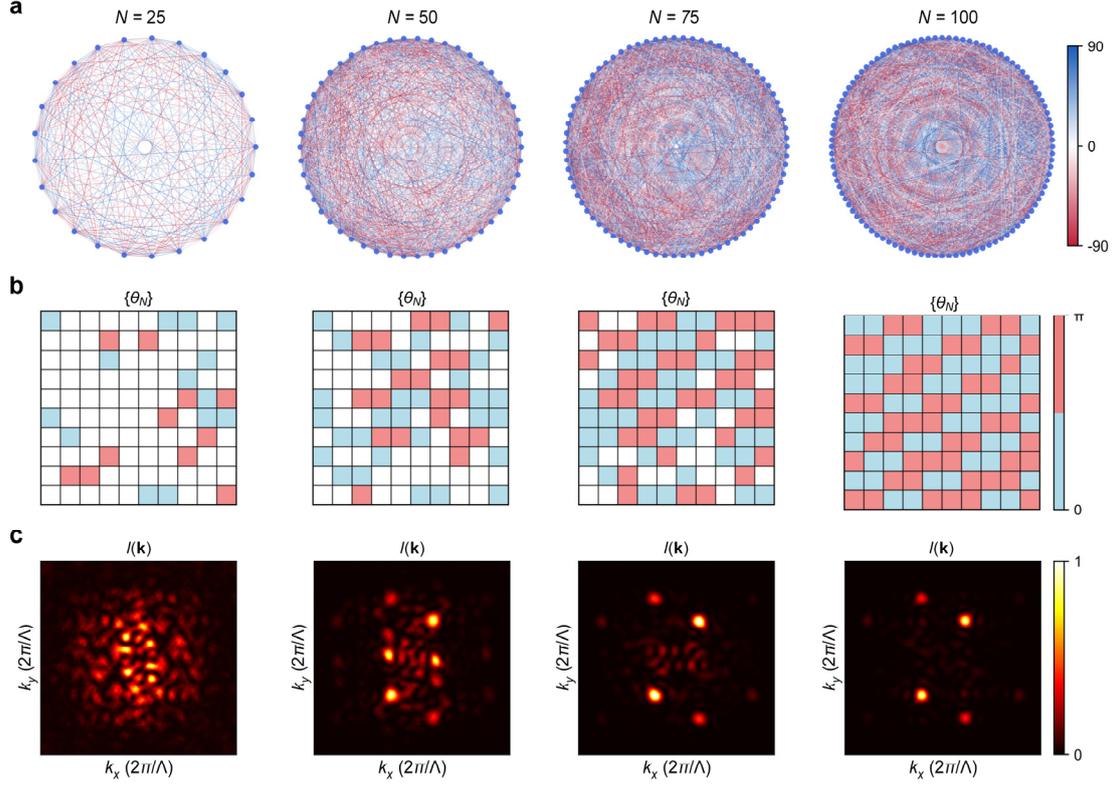

**Fig. 5 | Scaling behavior of the SPIM in real-space and k-space above the ET bound**. **a**, All-to-all connected networks with spin numbers $N = 25, 50, 75$, and $100$ on a $10 \times 10$ lattice. **b**, Experimentally obtained ground states manifested phase configurations ($0$ and $\pi$) in real space. **c**, Corresponding evolution of ground states intensity distributions $I(\mathbf{k})$ in **k**-space with increased spin number.

We further verified our system on a lattice of $60 \times 60$, with the number of spins gradually filling to $N = 300, 600, 900$, and $1200$. The graph parameters, experimental results, and performance metrics are summarized in Table 1. In the largest case, the spin number extends to $N = 1200$, yielding 719,400 connect edges, and the coupling matrix $\{J_{ij}\}$ containing 7,038 unique integer values. Notably, the experimental Hamiltonian $H_\mathbf{k}$ exhibited exceptionally high accuracy ($R^2 > 0.99$) compared with the results of Fig. 4d. This robustness arises from the fact that the coupling weights follow $J_{ij} = J(\mathbf{r}_{ij})$, so that multiple identical displacement vectors $\mathbf{r}_{ij}$ share the same coupling weight. The intrinsic path-overlap mechanism thereby averages out local perturbations and enhances noise tolerance. Due to the previously mentioned localization of the optical field (Fig. 5c, rightest panel), as the system approaches the ground state, the optical field gradually concentrates onto only a few pixels on the camera. This exceeds the detector's linear dynamic range, causing signal distortion and reducing the solution accuracy to ~88% for the $N = 1200$ instance.

Table 1 A summary of graph parameters and experimental performance.

| | 10 ×10 Lattice | | | | 60 ×60 Lattice | | | |
|---|---|---|---|---|---|---|---|---|
| Spins ($N$) | 25 | 50 | 75 | 100 | 300 | 600 | 900 | 1200 |
| Coupling resolution | 139 | 170 | 175 | 180 | 6198 | 6800 | 6969 | 7038 |
| Number of edges | 300 | 1225 | 2775 | 4950 | 44850 | 179700 | 404550 | 719400 |
| Average $R^2$ | 0.980 | 0.994 | 0.997 | 0.992 | 0.990 | 0.996 | 0.996 | 0.994 |
| Solution accuracy (%) | 100 | 100 | 100 | 100 | 98.11 | 94.16 | 92.12 | 88.48 |

**Discussion**

Previously, the coupling resolution of each Ising machine platform is fundamentally constrained by distinct physical mechanisms—for example, CMOS systems are limited by the discrete values of available capacitors and resistors, Mach-Zenhder interferometers by the nonlinear voltage-to-phase response of individual arms, coherent Ising machines by the memory bandwidth of programmable gate arrays, and SPIMs by the finite phase resolution of SLMs (8-bit). More importantly, it is critical to ensure that the corresponding energy or phase control signals remain robust against system noise —including detection noise, thermal fluctuations, and quantization errors— throughout propagation and computation. We summarize the coupling resolution of various physical Ising machines in Fig. 6, with all data points extracted from explicitly demonstrated problems in these literatures. As shown, the majority of existing architectures only support 2-bit couplings. In contrast, our ET SPIM provides 7,038 (res star) levels of fine couplings ($R^2 = 0.994$). This performance does not yet reach the intrinsic limit of the modulation function $V(\mathbf{k})$, but is instead constrained by the linear dynamic range of our detector (QHY294M, 13.6 stops). This limitation can be effectively resolved by employing a high-dynamic-range camera, such as the Photon-focus MV-D1024 with 20 stops[52].

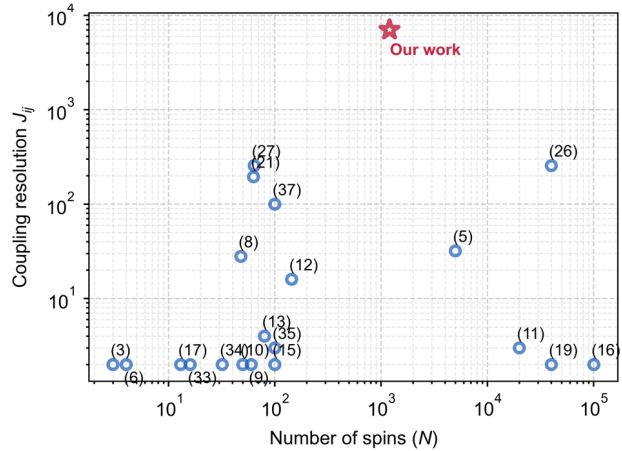

**Fig. 6 | Comparison of the experimentally demonstrated coupling resolution across different Ising machines**. Representative platforms are categorized into quantum systems (trapped ions *(3)*, single photons *(6)*, D-Wave *(5)*), electronic circuits (memristors *(9)*, coupled ring oscillators *(8)*), coherent Ising machines (CIMs) (Degenerate optical parametric oscillator *(16, 26)*, injection-locked *(17)* and opto-electronic oscillators *(15)*), CMOS annealers *(11, 12)*, photonic accelerators MZI *(21)*, SPIMs *(19, 27, 33, 34, 35, 37)*. surface acoustic waves *(10)*, superparamagnetic tunnel

junctions *(13)*. Our achieved coupling resolution is marked by a red star.

Above the ET bound, the coupling formats (Fig. 2c and Fig. 2d) naturally correspond to the generative dynamics observed in many real-world complex networks[46], including microbiome networks, social influence networks[53], epidemic spreading models[54], brain connectomes[55], and supply chain systems[56]. For instance, in microbiome networks, each spin represents a microbial species, while the coupling term $J_{ij}$ encodes their ecological interaction (e.g., promotion or inhibition). The introduction of repeated couplings, and creation of novel ones, respectively corresponds to "conservative evolution" and "innovative evolution," thereby modeling dynamic ecological phenomena such as species invasion, environmental shifts, and community reorganization[57,58]. Moreover, the framework of our ET SPIMs can be naturally extended to encode spin models with arbitrary external field[59] without sacrificing the scale of spin number (S6 for details).

Conventional digital computers exhibit quadratic scaling in both time and energy consumption with problem size $N$ when performing matrix multiply–accumulate (MAC) operations. In our approach, the computation time remains independent of the spin number $N$. This invariance arises because the pixel dimensions of $V(\mathbf{k})$ and $I(\mathbf{k})$ are fixed, while both the SLM modulation time and the camera acquisition time remain constant. The dominant time overhead in our current system stems from the SLM refresh rate (60 Hz), which can be increased to 20 kHz by replacing it with a digital micromirror device[20,36]. Moreover, $V(\mathbf{k})$ can be implemented using alternative optically selective hardware, such as amplitude modulators (Digital Micromirror Devices) or meta-surfaces, and combined with a photodiode to replace the camera, thereby performing all MAC operations entirely within the photonic domain. We estimate that when $N = 864$, our ET SPIM will surpass digital computers in solving time (S7 for details), in consistent with previous discussions[34,60]. In our work, the optical illumination power measured at the camera was only 25 nW, with this photon-domain energy consumption remaining independent of the problem size. By contrast, the most energy-efficient Ising machines based on memristors and stochastic electronic oscillators[13] exhibit energy consumption that increases substantially with system size.

**Methods**

Fig 1c illustrates the schematic of our experimental setup. A collimated Gaussian beam is generated from a 532 nm supercontinuum laser source. The beam waist is expanded 15-fold using two pairs of lenses to produce a uniform amplitude profile across the beam cross-section. This uniform beam is directed onto a phase-only SLM (CasMicrostar FSLM-2K39-P) for phase encoding. A Fourier lens (focal length $f = 150$ mm) performs the Fourier transformation of the modulated optical field. A 14-bit CMOS camera (Basler QHY294 Pro) is positioned at the Fourier plane to capture the optical interference pattern. The effective detection area is defined as 970 × 970 pixels, corresponding to the physical extent of the measured intensity distribution $I(\mathbf{k})$. The total optical power impinging on the camera is measured to be 25 nW.

**Supplementary Materials**

This PDF file includes:

**Acknowledgments**

B. W is supported by National Key Research and Development Program of China (2022YFA1205101), National Science Foundation of China (12274296, 12192252), Shanghai International Cooperation Program for Science and Technology (22520714300), Shanghai Jiao Tong University 2030 Initiative, and Yangyang Development Fund. E.H. acknowledges financial support from the Israel Science Foundation (grant number 1170/20).

**Data Availability**

The data are available from the corresponding author upon reasonable request.

**Competing interests**

The authors declare no competing interests.

**Author contributions**

B.W, H.L and G. W conceived the idea. E.H, B.W and X.C supervised this work. H. L performed the experiments, algorithm and partly theory. G.W conducted the theory and analysis. B.W and H. L wrote the manuscript. H. L and G. W prepared the supporting materials. H. L and G. W contributed equally to this work.